# Physical Light as a Metaphor for Inner Light

Liane Gabora
University of British Columbia

liane.gabora@ubc.ca

CONTENTS

Abstract

The metaphor between physical light and inner light has a long history that permeates diverse languages and cultures. This paper outlines a system for using basic principles from optics to visually represent psychological states and processes, such as ideation, enlightenment, mindfulness, and fragmentation versus integrity, as well as situations that occur between people involving phenomena such as honest versus deceptive communication, and understanding versus misunderstanding. The paper summarizes two ongoing projects based on this system: The 'Light and Enlightenment" art installation project, and the Soultracker virtual reality project. These projects enable people to depict their inner lives and external worlds including situations and relationships with others, both as they are and as they could be, and explore alternative paths for navigating challenges and living to their fullest potential. The projects aim to be of clinical value as therapeutic tools, as well as of pedagogical value by providing a concrete language for depicting aspects of human nature that can otherwise seem elusive and intangible.

## Introduction

This paper concerns what may well be the oldest, and to me the most fascinating, metaphor in all of human history: the metaphor between external light and inner light. To some, "inner light" has connotations of enlightenment, while others think of the creative spark, and still others use it to refer to life force, or chi. This paper will show how, by using external light as a metaphor to depict the often elusive workings of our inner lives it is possible to differentiate between these aspects of what is meant by inner light in a clear and tangible way, and gain a more concrete understanding of ourselves and each other.

   The paper begins with a brief discussion of the history of the metaphor. Then it outlines a visual language for using light to represent psychological and spiritual states. Finally it summarizes two projects aimed at putting this metaphor to work: an interactive art installation, and a virtual reality software program. Both enable users to visualize and creatively experiment with light-based representations of people in order to enhance understanding of self and others, to gain perspective on emotionally charged situations, and as a tool for creative self-expression. The approach appeals to our highly visual nature. If a picture is worth a thousand words, an immersive, interactive 3D animated world is worth many more. The metaphor derives its power not through persuasive arguments that touch the mind, but through imagery that works at an intuitive gut level.

   This program arises as part of a larger conceptual framework that seeks to capitalize on connections and resonances between what have been conceived of as different bodies of knowledge. Nature does not come divided up into subjects (such as physics, biology, psychology…); it comes as an unbroken whole. Although useful as an initial way of organizing knowledge, such categories are too often taken as reality, or as the way the human mind has evolved to carve up reality. Neither is the case. Phenomena such as the cross-domain recognition of personal style (Gabora, O'Connor & Ranjan, 2012; Ranjan, 2014), synesthesia (Ginsberg,1923; Ramachandran & Hubbard, 2001), the artistic practice of ekphrastic expression (wherein an artist attempts to have a more direct impact on an audience by translating the essence or form of a work of art from one medium to another), and indeed the mere fact that the movie-goer's experience is enhanced by a musical score, are evidence that nature, and our understanding of it, transcend domain-specific boundaries. The increasing acceptance of interdisciplinary research reflects a growing recognition that important questions, answers, and approaches fall through the cracks between what have been conceived of as different disciplines, leaving low-hanging fruit in the form of ideas with potentially economically viable applications for those who traverse disciplinary boundaries. In so doing, our understanding of reality grows richer, and our worldviews evolve; indeed this research direction grew out of a theory of creativity according to which it is worldviews, not memes, that evolve through culture, and creative thinking is what fuels their evolution (Carbert, Gabora, Schwartz, & Ranjan, 2014; Chrusch & Gabora, 2014; Gabora, 2000, 2005, 2013). The metaphor between physical light and inner light unites physics and psychology in the service of achieving a richer understanding of human nature, with potential benefits for public health, social harmony, and creative wellbeing.

### The History of Light-based Representations of the Psyche

The association between light and inner states of being has a long history. Religious history is replete with accounts of something not just vaguely light-like but an experience of a rarefied light that is felt rather than seen, and seems to burn from within. Eskimo shamen called it qaumaneq. Vedanta Hinduists call it Atma. The Tibetan Book of the Dead refers to it as the clear light of Buddha-nature. According to the Buddhist allegory of Indra's Net, humanity consists of a web made of threads of light stretching horizontally through space and vertically through time. At every intersection dwells an individual, and within every individual lies a crystal bead of light.

The metaphor between light and the essence of an individual is deeply woven into the human psyche. Light has been used as a metaphor for expressing vivid experiences of joy, spiritual insight, and creativity since the 'dawn' of civilization (Zajonc, 1993). It permeates our language, as in: enlightenment, moment of illumination, he beamed, her face lit up, to glow with enthusiasm, flash of brilliance, ray of hope, dim-witted person, light of my life, show me the light, dark night of the soul, shadowy nature, and so forth. Even cartoons have this property: everyone knows what it means when a light bulb appears above Charlie Brown's head. Organic processes, including cognitive processes, originate with and are made possible through the harnessing of light through photosynthesis (Wolken, 1986), and it has become cliché to say we are made of stardust. Thus it is not just in a metaphorical sense that we are beings of light.

Today, visualization is used to facilitate understanding widely across the sciences, social sciences, and popular culture as a means of conveying information about everything from weather patterns to stock market trends. However, its use to understand states of mind is limited. This is unfortunate, because although psychological, spiritual, and social phenomena often seem elusive and intangible, they are of unparalleled importance. Our lack of understanding of these arenas can lead to an overemphasis on the more straightforwardly understandable material aspects of life including external appearances. More importantly, miscommunication and lack of understanding of self and others may well be our greatest impediments to a sustainable, flourishing, happy future.

## Using External Light to Represent Inner Light: The Basics

We now look at how principles of optics (as outlined in texts such as Holtmannspötter & Reuscher, 2009; Reinhard, Khan, Akyuz & Johnson, 2008; Valberg, 2005) provide an elegant and intuitive metaphor that renders thoughts, motives, and interactions visible and therefore able to be dealt with more concretely.

### Representing Life Force or Chi

The basic system for representing the human psyche is as follows. We begin by noting that a spherical structure with a higher refractive index than the surrounding medium traps and amplifies light, as illustrated in **Figure 1a**. The sphere represents the body, and the light represents the individual's creative life force or chi, which is both enabled and constrained by the physical body. The colour of the sphere reflects both what is inside and what is at the surface. Thus a sphere may appear dark because it actually *is* dark or because it has a shell around it that obscures its light. Someone who is sick and has little life energy might be represented by **Figure 1b**, whereas someone who is aloof, or pre-occupied, or who hides their true nature with a protective mask, might be represented as having a thick or opaque shell around their sphere, as in **Figure 1c**. The shell reflects more light back into the sphere, locally amplifying and trapping it.

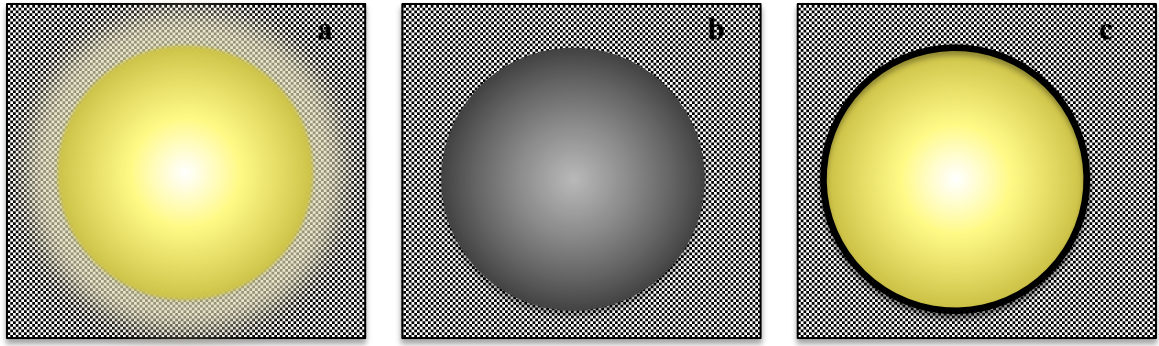

Figure 1. A cross-section of a spherical structure that has a higher refractive index than the surrounding medium traps and amplifies light (left). Such a sphere could appear dark from the outside either because it actually is dark inside (centre), or because its light is hidden (right).

**Reflecting on an Idea**
The metaphor between physical light and inner light can also be used to depict the subtle ways in which thoughts and ideas change as one thinks about them. Since the surface is concave, a ray of light diverges, or becomes less focused, as it passes through a sphere as illustrated in **Figure 2**. This can represent the stage an idea is at when you are able to think about it but not yet able to articulate it (i.e., put it into words) without distorting it.

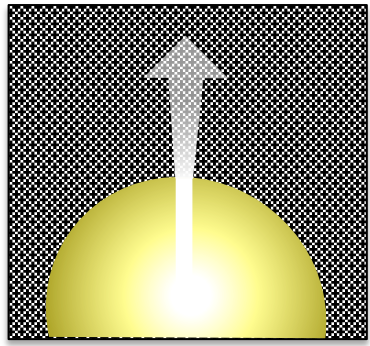

Figure 2. Since the surface is concave, a ray of light diverges, or becomes less focused, after it passes through the sphere. The more focused it is, the less it diverges.

However, if you think about an idea long enough you eventually do figure out a way of expressing it, and this too can be understood in terms of the metaphor. When an incident ray meets the surface of the sphere, it breaks into two: a reflected ray (which bounces off the inner surface of the sphere) and the refracted ray (which passes through the sphere), as illustrated in **Figure 3**. The refracted ray represents the externally detectable sense of focused concentration people exude when they are reflecting on a problem. Since the surface is concave, the ray becomes more focused each time it reflects. This can be used to represent how, as you "reflect" on an idea, or "bounce it around in your mind", it becomes more focused.

When a ray of light reflects off the interior surfaces of a sphere it reaches equilibrium such that it is no longer a single, distinct ray but now diffusely lights up the interior as a whole. This can represent how, once an idea has been reflected upon from all relevant perspectives it becomes so woven into the fabric of ones' worldview that it can be expressed in different ways, to different kinds of audiences, in a clear and focused manner.

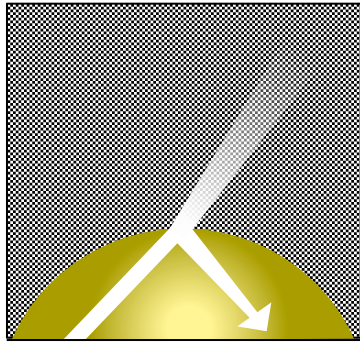

Figure 3. An incident ray (lower left) hits the interior surface of a sphere whereupon it breaks into two: a reflected ray (which goes back into the sphere), and a refracted ray (which passes through the sphere). Since the surface is concave, the refracted ray diverges and becomes less focused as it passes through the sphere, while the reflected ray converges and becomes more focused. Since the refractive index of the sphere is higher than that of air, more of the ray reflects than refracts. Since the surface is concave, the ray becomes more focused each time it reflects.

**Deep and Superficial Ideas**
When the point of origin of a ray of light is close to the interior surface of the sphere, there is only a tiny portion of the sphere through which it can project without significant refraction, because everywhere else it arrives at the surface of the sphere at an angle that deviates significantly from the perpendicular. Rays that originate close to the surface represent superficial thoughts (such as that you like someone's hair style), which are generally specific to a certain situation.

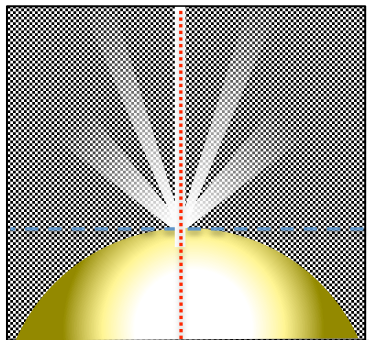

Figure 4. Light that originates near the edge of a sphere can only radiate in one direction; otherwise there is refraction and loss of intensity because it is not perpendicular to the sphere.

In contrast, when the point of origin of a focused ray of light is at the centre of the sphere, it can project in any direction without significant refraction, because it always arrives at the sphere at an angle that is perpendicular to the surface, as in **Figure 5**. Focused rays that originate close to the centre of the sphere represent deep thoughts (such as the concept of equality). They are generally relevant or applicable to many different aspects of life.

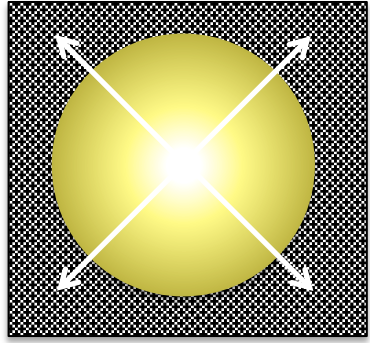

Figure 5. Light coming from the center of a sphere can radiate outward in any direction with no refraction and minimal reflection because wherever it contacts the sphere it is perpendicular to it.

**Communication**

Not just thinking, but also communication of information between people is represented by beams of light. When someone does not understand what has been communicated, this is represented as a ray of light bouncing off the surface of the sphere, as in **Figure 5a**. Misunderstanding can have different causes. It may be because the individual does not want to understand (as when someone is not interested). This is represented by an impenetrable shell surrounding the receiver of the information. Alternatively, misunderstanding can occur because the information was not put into a form in which it could be understood by the receiver (as when complex ideas are spoken to a child). This is represented by a beam of light that is not directed at the receiving sphere at a perpendicular angle, causing distortion and refraction as explained above. Another possibility is that the message is only partially understood. For example, it might be assimilated and responded to from a superficial layer of the self, as shown in the middle sphere. This is represented by an impenetrable shell within the sphere as in **Figure 5b**. Individuals may get into the habit of responding to each other from superficial layers of themselves, and come to direct their communication to each other *at* these superficial layers. The model suggests that when this happens, messages between them reliably undergo distortion, as represented by the fact that incoming and outgoing beams refract somewhat as they contact the surface of the sphere. If the message is communicated properly and understood, the message and response are represented without refraction, as in **Figure 5c**.

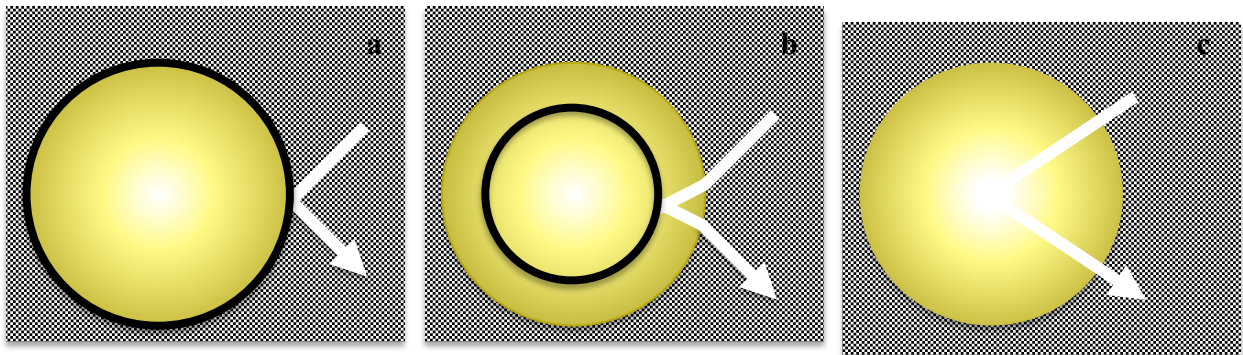

Figure 4. On the left a beam of light reflects off the surface of the sphere since it cannot penetrate it. In the middle, it reflects off a light-reflective surface without the sphere and there is refraction at the surface as it enters and leaves the sphere. On the right one beam is directed at the core of the sphere and another leaves the core of the sphere.

Communication of a clear or precise idea is represented using a sharp, focused beam. Such a beam may pass easily from one sphere to another even if it is not perfectly positioned. This represents that a straightforward message can be communicated even if one doesn't find the perfect words to express it. A vague or half-baked *idea* can be represented as a diffuse, defocused beam. Such a beam may require extensive reflection before it can be communicated effectively.

**Deception**
It has been shown that the proclivity to deceive others is highly correlated with a distorted perception of reality (Beck *et al*., 1990), and this phenomenon can be understood using the metaphor. A fracture (or vein of a different material) will cause a beam of light traveling through the sphere to bend (refract), and change direction. Lying, i.e., bending the truth, is thus represented as the deliberate use of a fracture to redirect a beam of light. When one is lying to someone else this happens at the surface of the sphere; if one is lying to oneself the refraction is occurring within the sphere.

Fracturing can represent, literally, a lack of integrity, a state wherein one is living with lies, or where one's values are not in sync with ones' actions, or one is living with memories that are too painful to face. Interestingly, the more fractured the interior, the longer it takes for a beam of light entering the sphere to reach equilibrium. Also, the greater the extent of fracturing, the less uniformly lit the interior will be when equilibrium is reached. There may be regions so fractured that light barely penetrates them. They represent the "shadow side" of the psyche, the aspects of oneself or one's life that one wants to avoid, as illustrated in **Figure 6**.

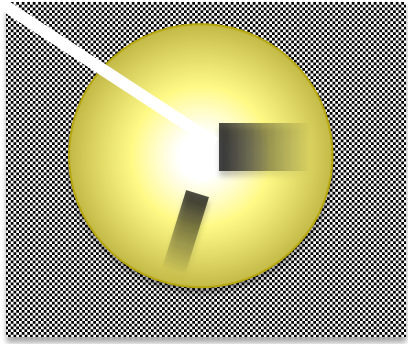

Figure 6. A beam of light coming from the upper left refracts (bends and changes direction), and becomes darker and less focused as it passes through a fracture, or may be unable to penetrate the fracture entirely. A fractured sphere may not be uniformly lit; it may contain shadowed regions that the light does not fully penetrate.

**Enlightenment**
When a ray of light reflects off the interior surfaces of a sphere it quickly reaches equilibrium such that it is no longer a single, distinct ray but now diffusely lights the interior. An individual who has achieved a state of enlightenment is represented as one who has no fragmentation or impurities in the psyche, such that the interior is rightly and uniformly lit, as in **Figure 1a.** Such an individual communicates from the core of the self as opposed to a superficial level of the self, because there is nothing blocking the core. Enlightenment in this model is not a rarified state; it is a state in which one is free of internal fragmentation and able to "be themself", achievable by almost anyone. It is proposed that mindfulness is the state of remaining alert to the presence of fragmentation or shadows and considering them from different perspectives to overcome them.

**Complex Thoughts and Feelings**
In addition to enabling the depiction of general properties of a psyche as we have seen above, the metaphor enables people to break thought patterns down into their constituent components by representing specific feelings, knowledge, values, and assumptions that make up a recurring pattern of thought as light of different colours and frequencies. Let us say, for example, that a compulsive desire to check that the door is locked is represented by the high-frequency pattern depicted in **figure 7a**. One might come to realize that there is something lurking behind that compulsion: a memory of being intruded upon. This is depicted in **b**, giving rise to the more complex pattern **c** that combines the two. **Figure 8** depicts the situation wherein an individual's response, **c**, might be greater than expected because it reflects a cumulative build-up of similar events **a** and **b**. **Figure 9** depicts the opposite situation, wherein an individual's response, **c**, might be less than expected because the event that might trigger this response, **a**, has been cancelled out or nullified by another event **b**.

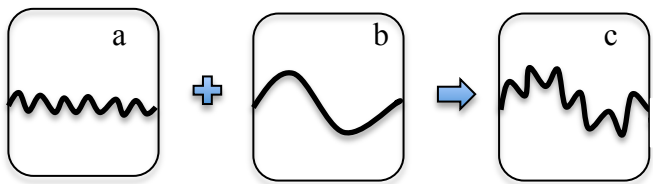

Figure 7. Wave c is the sum of the superposed waves a and b.

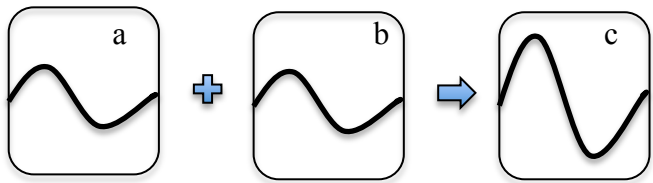

Figure 8. The amplitude of wave **c is equal to the sum of its component waves a and b** due to constructive interference.

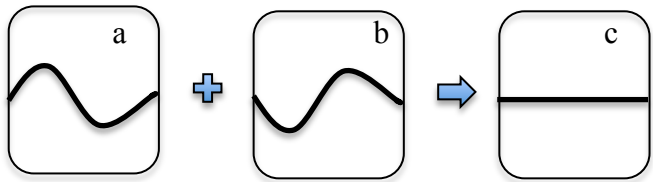

Figure 9. Waves **a** and **b** exhibit destructive interference such that the result is no wave at **c.**

Thus, using the properties of light as a metaphor one can model how patterns of thought and behavioral responses are made up of components, which can be addressed one by one.

## Applications

We have only scratched the surface of the metaphorical representational system but that is the basic idea. It aims to have both therapeutic value as a form of art therapy wherein the artistic outcomes have relatively straightforwardly interpretable symbolic meanings, and as a tool for doing psychological research for gaining insight into human nature. The research component is set directly and actively *within* a sustained creative process of generating and exploring possibilities through light-based representations.

Static images of light are of limited use as a metaphor for understanding real human situations, which are dynamic, and which could unfold different ways in different conditions. The next two sections of the paper outline attempts to realize the dynamic potential of the metaphor, first in an art installation, and second in a piece of software.

### Light and Enlightenment: An Interactive Art Installation

The 'Light and Enlightenment' project, still in its infancy, is an aesthetically appealing and inspirational interactive art installation that enables users to visualize and creatively experiment with light-based representations of themselves and others, thereby clarifying feelings and events going on in their lives. In the tradition of artists such as Dan Flavin, Bruce Naumen, or James Turrell, light is used not just to illuminate something else; it is an intrinsic part of the artistic conception. In the sense of light acting as a representation of a conscious being, the approach bears some relationship to Popat and Palmer's (2009) work with people interacting and connecting with light sprites through dance. The project aims to exert a meaningful impact on how people explore and develop understanding of the ways people are interconnected, and of the relationship between life events and artistic ideas.

The allegory of Indra's Net (mentioned above), and other uses of the light metaphor mentioned in the Introduction, will be written on a display at the entrance. Extending around the perimeter of a dimly lit, mirrored room will be a slowly undulating web of aluminum tubing covered with EL-wire, guided at the edges by rollers. At every vertex is a resin sphere containing a candle. Candles can be lit by attendees, and put into a spot from which they are drawn into the web. The goal of this first chamber of the installation is to remind attendees that we are all connected, and to have this be alive in their memories as they explore situations of potential isolation or conflict.

In the main room of the art installation, a person's creative life force is represented as light, and their internal model of reality, or *worldview,* is portrayed as a spherical entity that amplifies and transforms this light. Attendees generate visual depictions of their inner workings by controlling how light moves through and between movable acrylic spheres of different sizes, colors, and degrees of transparency, hanging 4 to 6 feet above ground. They represent the hidden dynamics of their inner lives and relationships using handheld devices and toggles that control the properties of the spheres, including their relative positions and how they interact. Some spheres are detachable, allowing for physical, embodied interaction with them, which is expected to enrich attendees' experiences (Antle & Corness, 2013). Some spheres are non-detachable because they are connected by servo motor links to projectors and a computer, and operated by toggles and handheld devices attached to control panels.

A toggle lets you vary the amount of light in a sphere to indicate how vibrant or alive the person is. Other toggles allow you to vary properties of the sphere such as the thickness, transparency, or colour of its outermost shell. The user is encouraged to use a thick shell to represent someone who is guarded and reserved, and a thin shell to represent someone who is open and friendly. There are toggles that allow the user to 'paint' different regions of the shell different colours, representing different arenas of life, or to create multiple spheres embedded in one another, i.e. 'layers of the self'.

Another set of toggles is used to represent thinking and communicating using beams of light. The user is invited to explore how varying the size, intensity, colour, diffuseness, and direction of the beam relative to the sphere(s) affects reflection and refraction at sphere interfaces, and shown how these parameters can be used to represent phenomena such as reflection on an idea, miscommunication, deep versus superficial ideas, and so forth, as described in the previous section. Another toggle lets the user depict attraction or creative resonance between people as sparks that appear to flicker between the two spheres.

Yet another set of toggles enables the user to represent phenomena such as deception and repression, by generating fractures, dents, and internal boundaries of various shapes, sizes, and degrees of transparency, that distort how light flows in a sphere. The user can represent an emotionally charged topic by creating an **opaque bubble** in the sphere. The fact that light has difficulty both entering the bubble and leaving the bubble represents that the person probably either avoids the topic, or dwells on it excessively (or both, at different times). The user is invited to experiment with how fragmentation generates shadowy regions, and how the presence of shadowy regions leaves a detectable trace in the overall appearance of the sphere and the beams of light that leave it.

**SoulTracker**
The SoulTracker is an immersive, interactive virtual reality for depicting and playing with representations of inner light of self and others. It works along the same principles as the art installation, and enables users to do everything they did there, and more. In addition to providing the capacity to model oneself and others using spheres and beams of light, it also provides:
- More ways to depict aspects of one's inner life and interactions with others. For example, the degree of comfort people feel in each other's presence can be indicated by how blurred or sharp-edged their borders are when they are depicted in close proximity.
- The ability to make a copy of any scenario and use it as a starting point for depicting how it looks from different perspectives, how it arose, and what might happen next.
- The ability to create cross-sections of spheres and thereby see inside opaque spheres.
- The ability to save visualizations, email them, print them, post them to the cloud, etc.

Thus SoulTracker facilitates awareness and exploration of the potential each moment holds for creatively reconceptualizing the present and putting possible future outcomes within reach.

A working prototype of the SoulTracker has been built. The screendump in **Figure 5** gives a simple example of how it is used to portray a hypothetical situation. Using *view mode* the user depicted two bonded individuals (fuzzy spheres), one of whom is deceiving the other (refracted beam), and a third (sharp-edged sphere in shadow) who is left out. In *overview mode* a miniature version of the situation would be shown, with events that led up to it to the left, and ways it could unfold to the right. The SoulTracker is still in a rudimentary state. Shadows and beams are not realistic, and the physics of light has not been built in; for example, reflection and refraction of light at surfaces and fractures must be drawn manually rather than happening automatically.

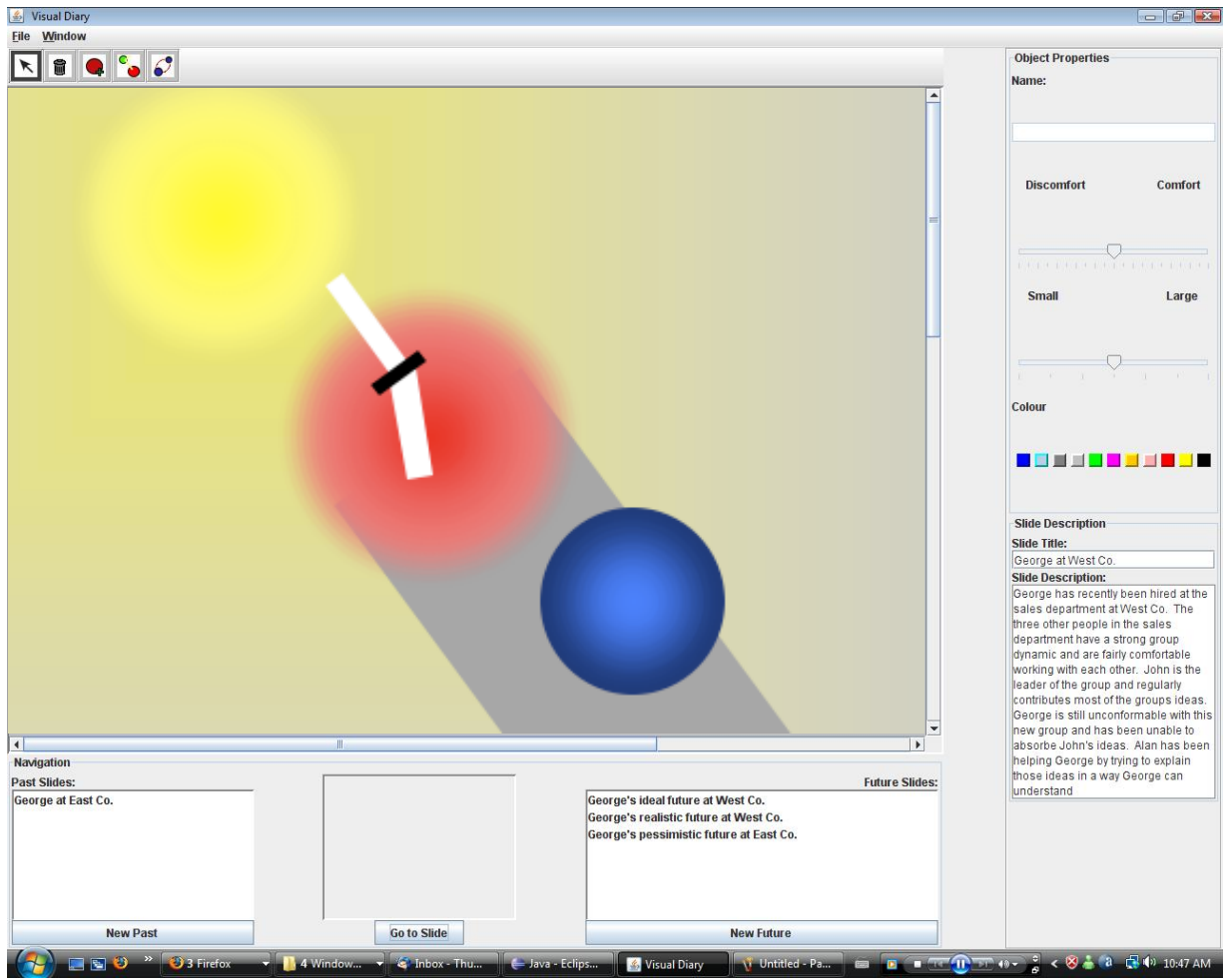

Figure 5. A screendump from SoulTracker (see text for details).

In the new version of SoulTracker, the physics of how light reflects and refracts off concave/convex surfaces and fractures, and how it is amplified, focused, diffused, or changes color due to various structural features of spheres will be built in. This will make it easier not just to portray life situations but to experiment with them and observe the consequences of modifying different aspects of the situations. For example, wrestling with a problem will now be represented as a beam of light that repeatedly reflects within the sphere, and to some degree refracts and thus escapes at the sphere's periphery, thereby subtly changes the ambient light and affecting others. Other new capacities will be added, such as the ability to create the appearance of slowing down the beams of light so that the user can watch the internal workings of spheres and their interactions. For example, it will be possible to observe the process by which a sphere regains equilibrium after receiving an incoming light beam (which represents assimilating new information), or direct light at fragmented regions until shadows disappear. The software will use a neural network to learn from user inputs to make suggestions for how depicted scenarios might unfold based on similar situations that have been entered before. It will be possible to view (and even modify) how a given scenario might look from the perspective of someone else by clicking the sphere that represents that person. Members of families, organizations, or business can share their different perspectives of the same situation, and thereby learn from each other. By posting their scenarios to the cloud users will be able to locate others who are in similar situations to their own, and show each other how they are coping and how they are feeling, i.e., communicate with them using this "language of light" even if they don't speak the same language.

**Testing, Development, and Expected Outcomes**
The work described here is at an early stage. Steve DiPaola, Maria Lantin, and I are currently seeking funds to work on the project and collaborators in computer graphics, optics, and digital technology. Our plans include not just building but testing the effectiveness of the technology.

Consenting users of the interactive art installation will be filmed as they create and work with visual depictions of their inner processes using the interactive art installation and the software. They will be interviewed about what insights they have gleaned about themselves, their relationships, their creative process, and their sense of purpose and self-understanding through the creation of these dynamical depictions. Exposure to the installation is expected to enhance users' understanding and control over their inner lives and interpersonal situations, and give evidence of breaking out of habitual patterns and approaching situations in new and more effective ways, as well as a sense of being part of something larger than oneself. Though the focus of this project is not on the external output but on the internal processes of understanding, creating, and portraying, the project is expected to generate art that is provocative and beautiful.

The effectiveness of SoulTracker as a therapeutic tool for artists will be investigated in experiments carried out over ten 90-minute sessions with consenting users. They will be divided into an experimental group that uses SoulTracker and a control group that watches entertaining videos. Both groups will be given a questionnaire to assess their degree of awareness and control over life problems and interactions with others, and capacity to achieve resolution and self-understanding through their creative process. Introduction to SoulTracker in the first session will be accompanied by discussion of the scientific and psychological framework. After that, and in subsequent sessions carried out at computers with a facilitator available to help and answer questions, they will be encouraged to depict and play with thoughts, creative ideas, situations and possible future developments of them. SoulTracker sessions from consenting users will be anonymously analyzed for evidence of (1) sense of control over and understanding of situations, (2) awareness of the present moment and its rich possibilities for creative decision making and breaking out of habitual modes of thinking and acting, (3) ability to communicate, take into account perspectives of others and respond with empathy, and (4) evidence of enhanced understanding or resolution of life events through creative re-interpretation of them. In addition, the questionnaire given in the first session will be re-administered to both groups to assess the therapeutic value of the SoulTracker. A second questionnaire will be administered to the experimental group only to determine how personally useful they found the SoulTracker.

Use of these tools is expected to facilitate (1) the realization that there are an infinite number of ways of constructing the tapestries of understanding from which our thoughts and actions emanate and thereby affect others, (2) the weaving of difficult to verbalize but emotionally charged situations into a form in which it is possible to comprehend, explore, and come to terms with them, and (3) creative problem solving. By better understanding how users are using the software and how their use changes across sessions we can gain a better understanding of its effectiveness and how to improve it. Analysis of the actual scenarios generated and accompanying written comments using the SoulTracker sessions is expected to provide evidence of self-discovery and of new ways of using the software to shed light on human nature.

The research program generates related avenues for further investigation. An important next step toward getting a better handle on what we actually mean by the term inner light. One means of accomplishing this involves assessing the extent to which there is agreement amongst people's assessments of the degree to which someone exudes (or obstructs) their inner light using a modified version of a research protocol that has been previously used to assessing the extent to which there is agreement amongst people's assessments of the degree of authenticity in creative performances (Henderson & Gabora, 2013).

**Theoretical Framework in which these Projects Reside**

The work presented here grew out of earlier applications of optics to model the evolution of worldviews and consciousness (Gabora, 1999, 2002). More broadly the research program arose as part of a scholarly effort to develop a scientific framework for cultural evolution based on the hypothesis that what evolves through culture is our internal models of the world, or *worldviews:* including stories, memories, knowledge, values, and beliefs, with both cognitive and emotional components (Gabora, 2000, 2004, 2008a, 2013). Significant progress has been made in modeling cultural evolution using a combination of computer modeling (Gabora, 2008b,c; Gabora & Leijnen, 2009; Leijnen & Gabora, 2010; Gabora, Chia, & Firouzi, 2013) and mathematical modeling (Aerts, Gabora, & Sozzo, 2013; Gabora & Aerts, 2009; Gabora & Kitto, 2013). This project has also advanced through empirical studies of human creativity (DiPaola & Gabora, 2007, 2009; Gabora, 2000, 2005, 2010; Gabora, O'Connor & Ranjan, 2012; Gabora & Saab, 2010; Gabora & Ranjan, 2013). After all, just as *biological* evolution is driven by the processes that generate adaptive novelty (e.g., recombination, mutation, self-organization, and epigenetic mechanisms), *cultural* evolution is driven by human creativity.

In order for this research to exert a significant impact on *how* culture actually evolves, it is necessary to not just study cultural change *from the outside* but provide opportunities for growth *on the inside.* Such opportunities for growth must be broadly intuitive and engaging while targeting 'movers and shakers' of cultural change. The art, science, and technology with which we human change and make sense of document, and the cultural trajectories that we thereby touch, are affected by how we weave narratives about everyday matters. Thus, work on creative processes on the continuum between *big-c creativity* (the processes of eminent creators) and *little-c creativity* (everyday problem solving and spontaneous creative behavior) can lead one to questions about what Beghetto and Kaufman (2007) refer to as *mini-c creativity*, which they define as the novel and personally meaningful interpretation of experiences, actions, and events. One can argue that mini-creative acts, because of their universality, exert at least as great an impact on the unfolding of human culture as the masterpieces of creative giants. The inner light program described here provides opportunities to visualize ones' worldview as part of an evolving tapestry of interacting worldviews. It prompts micro-moments of reflection on one's ways of being and relating, and could therefore affect the myriad thoughts and acts that together constitute human cultural evolution.

**Summary and Conclusions**

This paper described a program of research that aims to have a meaningful impact on how people explore and develop ways of understanding themselves, their creative process and its relationship to life events, and their relationships to others and to their community. There are several components, unified by an underlying metaphor between physical light and inner light, which can refer to creative spark, life force, or spiritual essence. Creative life force is portrayed as ambient light, and people's psyches are represented by spheres that amplify and transform light. We have seen how personality characteristics, situations, and relationships can be systematically depicted using a systematic visual language based on the properties of light and how it interacts with physical objects. For example, vibrant people are portrayed as having lots of light, and aloof people as having spheres with thick shells. Thoughts and ideas are represented as beams of light that converge—become more focused—when reflected off the concave inner surface of a sphere, and refract (bend) when they are (intentionally or unintentionally) misunderstood. Vague ideas are represented with diffuse beams that require much reflection. Thus the metaphor turns elusive aspects of human nature, and the situations we find ourselves in, into concrete visual depictions that can be explored and experimented with. This paper does not go into extensive detail about how the metaphor can be used but hopefully there is enough information to convey the basic idea and tantalize the imagination.

    The art installation enables attendees to visualize and creatively experiment with light-based representations of themselves and others. Handheld devices attached to control panels enable attendees to direct beams of light of different sizes and intensities through and between spheres. Toggles enable attendees to control qualities of a sphere such as its colour, opacity, and level of ambient light, and the user can generate regions of fragmentation that create shadows and distort the flow of light within a sphere.

    The SoulTracker software works along the same principles as the art installation, but offers a more private forum for visually depicting peoples' inner processes using spheres and beams of light, and provides some extra features. For example, the degree of comfort people feel in each other's presence can be indicated by how blurred or sharp-edged their borders are when they are in close proximity. It also allows users to save depicted scenarios, and to use them as a starting point for depicting how the same situation looks from different perspectives, or how it arose, or to create different possible 'future scenarios' for the depicted situation.

    In short, research on inner light is useful for visualizing and understanding communication and miscommunication, wholeness and fragmentation, honesty and dishonesty, closeness and isolation, potentiality and actualization, and the process by which creative ideas are born and take shape. The program described here unites the *arts* (installation art) with the *natural sciences* (optics; digital technology) and the *human sciences* (cognitive, social, humanistic psychology) in a robust framework. It is hoped that providing a means to visually depict the intangible but all-important psychological and spiritual elements of human existence using a straightforward "language of light' will facilitate the evolution of worldviews that are integrated, adaptive, and humane.

## Acknowledgements
This research was conducted with the assistance of grants from the National Science and Engineering Research Council of Canada, and the Fund for Scientific Research of Flanders, Belgium. Thanks to Camille Selhorst for helpful comments on an initial draft.